\documentclass[reprint,twocolumn,aps,showpacs,floatfix,amsmath]{revtex4-1}

\usepackage[english]{babel}
\usepackage{graphicx}
\usepackage{epsfig}
\usepackage{color}
\usepackage{amssymb}
\usepackage{amsmath}
\usepackage{array}
\usepackage[loose]{units}
\usepackage[pdftex]{hyperref}
\hypersetup{breaklinks=true}
\usepackage{braket}%
\usepackage[caption=false]{subfig}
\usepackage{cleveref}
\usepackage{appendix}

\usepackage{letltxmacro}

\LetLtxMacro{\ORIGselectlanguage}{\selectlanguage}
\makeatletter
\DeclareRobustCommand{\selectlanguage}[1]{%
    \@ifundefined{alias@\string#1}
      {\ORIGselectlanguage{#1}}
      {\begingroup\edef\x{\endgroup
         \noexpand\ORIGselectlanguage{\@nameuse{alias@#1}}}\x}%
}
\newcommand{\definelanguagealias}[2]{%
  \@namedef{alias@#1}{#2}%
}
\makeatother

\definelanguagealias{en}{english}

\newcommand{\UO}{Department of Physics, 1274 University of Oregon, Eugene, Oregon, 97403}

\newcommand{\deriv}[2]{\frac{d #1}{d #2}}

\begin{document}

\title{Demonstration of electron helical dichroism as a local probe of chirality}
\author{Tyler R. Harvey}
\author{Jordan S. Pierce} 
\author{Jordan J. Chess}
\author{Benjamin J. McMorran}
\email{Correspondence may be addressed to mcmorran@uoregon.edu}
\affiliation{\UO}

\date{\today}

                            
\begin{abstract}
  We report observation of electron helical dichroism on a material with chiral structure. In analogy with circular dichroism, a common technique for molecular structural fingerprinting, we use a nanofabricated forked diffraction grating to prepare electron vortex beams with opposite orbital angular momenta incident upon metal nanoparticle clusters and post-select for a zero-orbital angular momentum final state. We observe a difference in the differential scattering probability for orbital angular momentum transfer from vortices with opposite handedness incident on chiral aluminum nanoparticle clusters at $\unit[3.5\pm0.8]{\textrm{eV}}$.  We suggest that the observed electron helical dichroism is due to excitation of surface plasmon vortices. Electron helical dichroism enables chirality measurement with unprecedented spatial resolution over a broad range of energies.


\end{abstract}

\maketitle

Chirality -- the absence of symmetry under parity -- lies at the heart of a variety of open research questions, including CP violation, the existence of an elementary majorana fermion, magnetic skyrmion behavior, and broken symmetry in the biochemistry of life. Chirality can also serve as an easy proxy for properties that are more difficult to directly measure. For example, sugar molecules consumed and produced by living organisms are always right-handed; to measure the concentration of a sugar in solution, one can measure the rotation of linearly polarized light passed through the solution \cite{mcnichols_optical_2000}. 

Light has long been used to measure the chirality of ensembles of molecules, and more recently, increasingly small engineered structures. Circular dichroism (CD) is a standard spectroscopic tool for structural fingerprinting of molecules through measurement of three-dimensional chirality associated with molecular structure at a particular length scale \cite{fasman_circular_1996}. Circular dichroism measures the difference in absorption of right- and left-circularly polarized light. Circularly polarized light carries chirality; the chirality of a massless particle is equivalent to its spin helicity,
\begin{equation}
  h_S = \mathbf{S} \cdot \mathbf{\hat{p}}
\end{equation}
where $\mathbf{S}$ is the spin angular momentum vector and $\mathbf{\hat{p}}$ is the propagation direction of incident light. Helicity, like chirality, is invariant under rotation and changes sign under parity. Chirality-sensitive interactions between circularly polarized light and engineered chiral structures have also recently been predicted theoretically \cite{prosvirnin_polarization_2005} and observed in experiments \cite{papakostas_optical_2003,kuwata-gonokami_giant_2005,ohno_study_2006,konishi_effect_2007}. Circular dichroism is sensitive to other kinds of chirality beyond material structure: X-ray magnetic circular dichroism (XMCD) characterizes the chirality associated with unpaired electron angular momentum in atoms that leads to magnetization \cite{schutz_absorption_1987}.

However, circular dichroism techniques can only survey limited energy ranges with diffraction-limted spatial resolution. Standard light sources for visible circular dichroism typically can efficiently produce only the lowest energy UV light, and atmospheric oxygen strongly absorbs UV over \unit[6]{eV}. However, UV circular dichroism spectra are necessary for amino acid \cite{meierhenrich_circular_2010} and protein secondary structural characterization \cite{johnson_protein_1990}; X-ray circular dichroism allows for local, element-specific magnetization determination in magnetic materials \cite{schutz_absorption_1987}. Far-UV and X-ray circular dichroism spectra are typically gathered at synchrotron light sources \cite{wallace_synchrotron_2011,schutz_absorption_1987}.

Electron microscopes, on the other hand, are equipped for spectroscopic analysis of materials over a stunning range of energies: a single electron microscope can quickly measure spectra across five orders of magnitude in energy. Monochromated, aberration-corrected electron microscopes can resolve excitations with energies as low as \unit[10]{meV} (far-infrared) and well into keV (hard X-ray) \emph{in combination} with sub-nanometer structural resolution on durable specimens \cite{krivanek_vibrational_2014}; good progress has been made towards atomic resolution on beam-sensitive materials \cite{zhou_towards_2008}. In this Letter, we demonstrate sensitivity to chirality in a new kind of spectroscopy with helical electron states.

Free electrons can be prepared in states with helical wavefronts, called electron vortices. Electron vortices are eigenstates of $L_z$, the orbital angular momentum operator along the direction of propagation $\langle \mathbf{\hat{p}} \rangle = \mathbf{\hat{z}}$. These states carry a property we call orbital helicity
\begin{equation}
h_{L} = \mathbf{L} \cdot \langle \mathbf{\hat{p}} \rangle = L_z. \label{eq:hL}
\end{equation}
The helicity of massive particles is a reference frame-dependent analogue of chirality: helicity is invariant under rotation, changes sign under parity, but also changes sign under a boost into a reference frame that reverses the direction of propagation. As seen in \eqref{eq:hL}, orbital helicity behaves differently under transformation but is equivalent to orbital angular momentum in a given reference frame. 

\begin{figure}[thb]
  \subfloat[]{\label{subfig:setup:grating}
    \includegraphics[width=0.57\columnwidth]{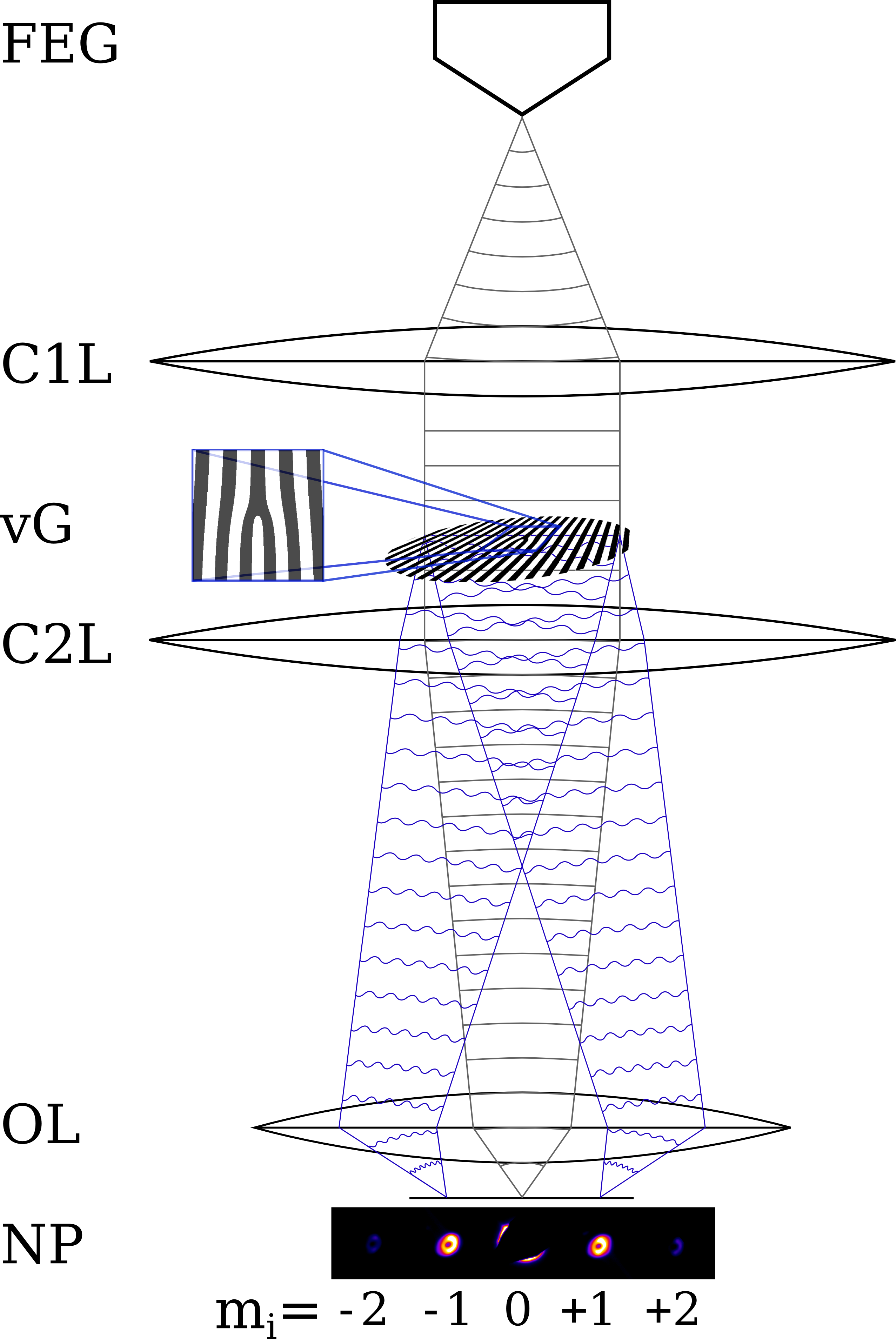} }
  \subfloat[]{\label{subfig:setup:beam}
    \includegraphics[width=0.38\columnwidth]{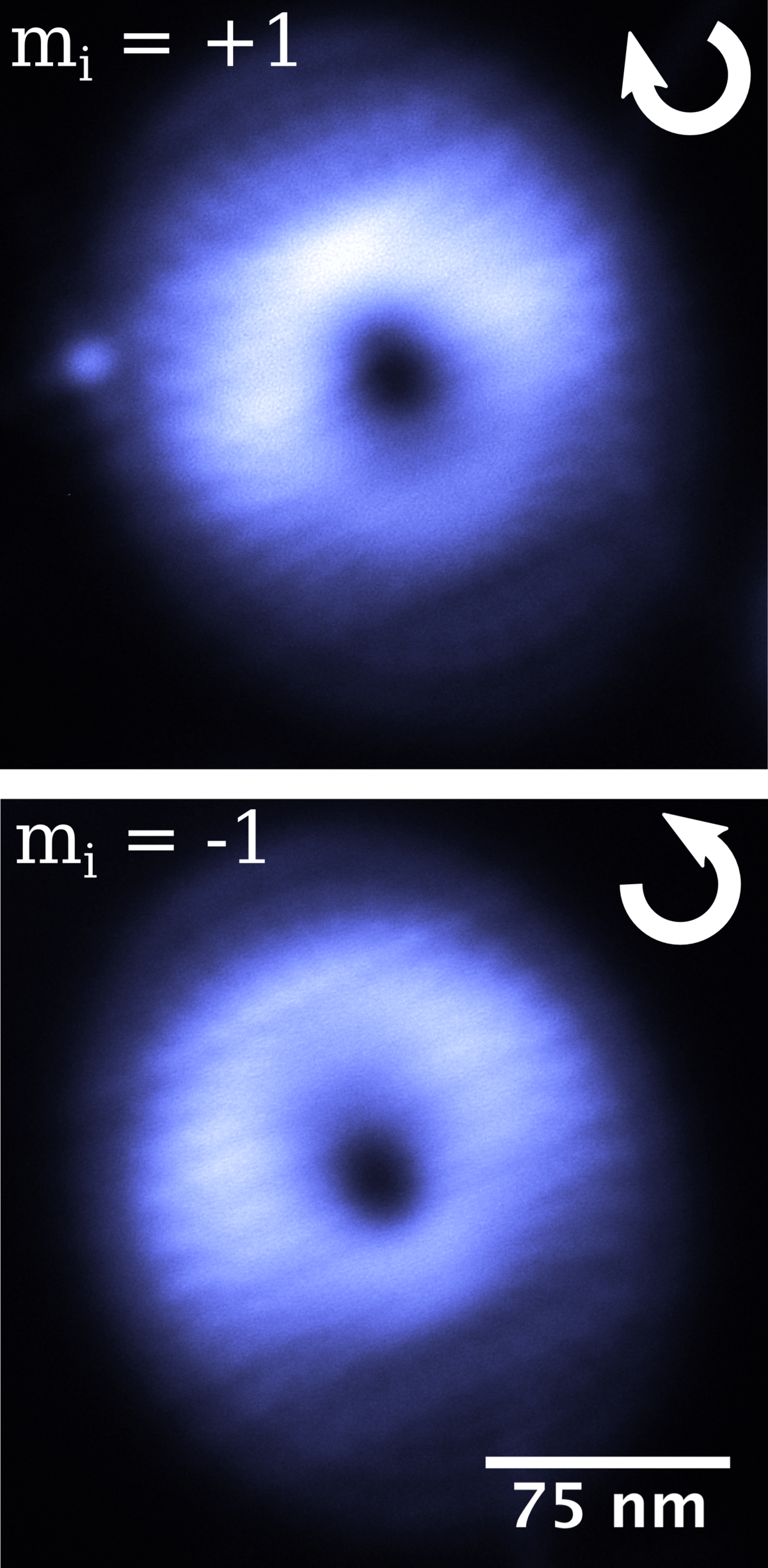} }
  \caption{
    (a) Schematic of probe-forming optics. Electrons produced by the field-emission gun (FEG) are accelerated through a gun lens (not pictured) to \unit[300]{keV}, and emerge from the upper condenser lens system (C1L) with planar wavefronts. The single-forked vortex diffraction grating (vG) adds an azimuthal phase and transverse momentum. The lower condenser lens system (C2L, OL) produces a focused far-field diffraction pattern at the specimen plane (NP); the $n$th diffraction order is a vortex with $m_i = n$ quanta of orbital helicity.
    (b) Transmission electron micrograph of an $m_i = -1$ (top) and $m_i = +1$ (bottom) electron vortices with nearly identical intensity distributions passed through silicon nitride substrate. Due to the phase singularity at the center of the beam, intensity is near zero there. 
   \label{fig:setup} }
\end{figure}

Electron eigenstates of orbital helicity can be prepared in a transmission electron microscope (TEM) by inclusion of an optical element that adds an azimuthal phase to the electron wavefunction \cite{uchida_generation_2010,verbeeck_production_2010,mcmorran_electron_2011}. The resultant wavefunction $\psi(\mathbf{x}) \propto e^{im\phi}$, where $\phi$ is the azimuthal angle and $m$ is the orbital helicity quantum number ($\langle h_L \rangle = m\hbar$). Nanofabricated diffraction gratings \cite{verbeeck_production_2010,mcmorran_electron_2011,grillo_highly_2014,harvey_efficient_2014}, material \cite{uchida_generation_2010} and magnetic phase plates \cite{beche_magnetic_2014,blackburn_vortex_2014} can produce beams with well-defined orbital angular momentum. The orbital helicity of electron vortices is well-suited for chirality measurement.

We report observation of electron helical dichroism on a material with chiral structure.  Electron helical dichroism is the electron-vortex analogue of circular dichroism. We record an electron energy-loss (EEL) spectrum for the interaction of incident single-helix electron vortices (with incident orbital helicity quantum number $m_i = \pm 1$) with a specimen, and we post-select for the component of the outgoing wave that has zero final orbital helicity ($m_f = 0$). When the probability of multiple-scattering is low, electron intensity as a function of energy lost is approximately proportional to a differential scattering cross-section (for volume excitations) or a differential scattering probability (for surface excitations) \cite{egerton_electron_2011}. An orbital helicity dependence in the electronic density of states appears as a peak in the electron helical dichroism spectra. Structural chirality of a specimen breaks the symmetry between positive- and negative-orbital helicity modes in the density of states. The dichroic interaction therefore probes the chirality of a specimen. Indeed, Asenjo-Garcia and Garc\'{i}a de Abajo recently predicted that dichroism is possible with electron vortices incident on a material with chiral structure \cite{asenjo-garcia_dichroism_2014}.

\begin{figure}[htb]
  \subfloat[]{\label{subfig:i-m:superposition}
    \includegraphics[width=0.53\columnwidth]{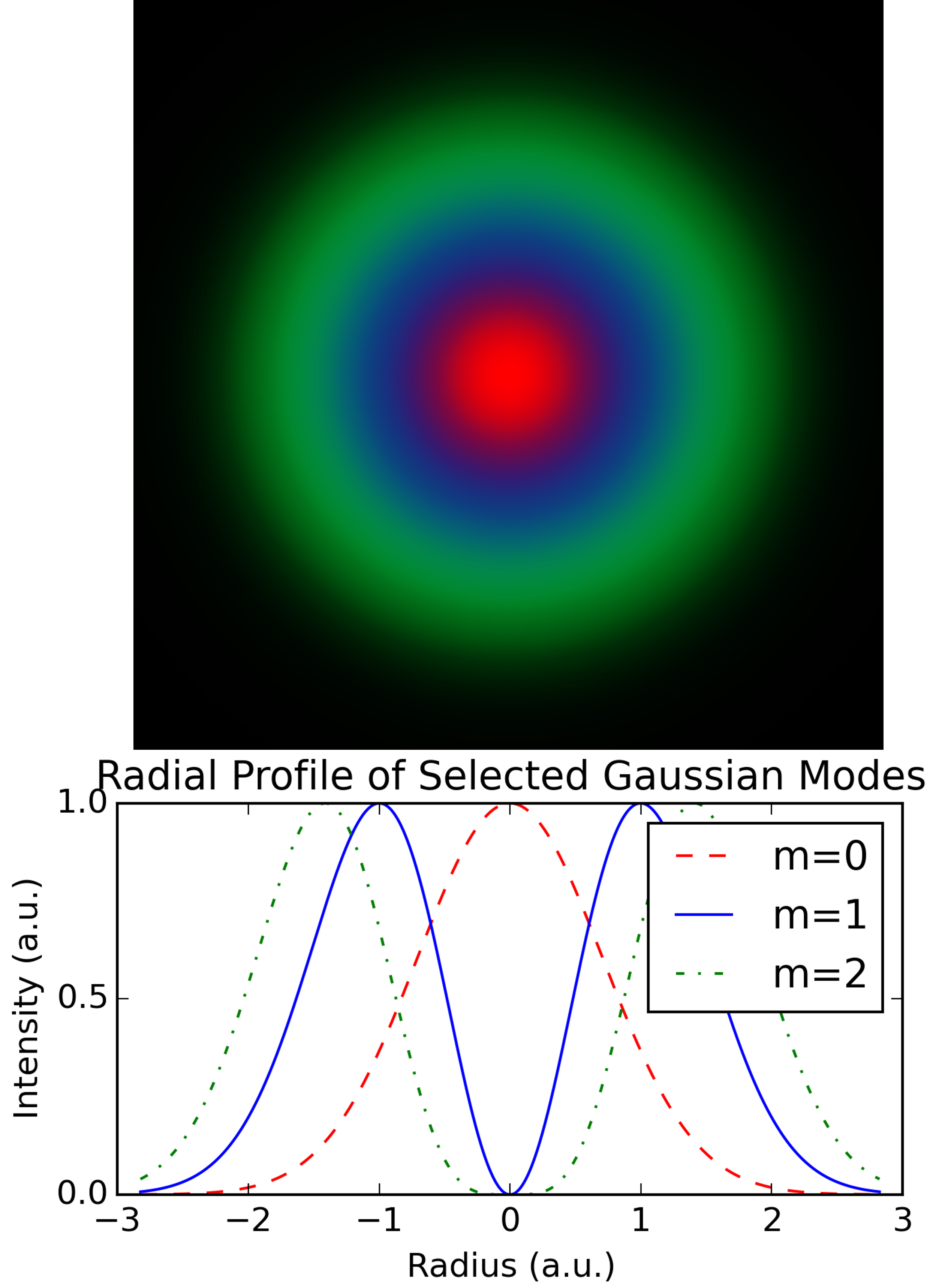} }
  \subfloat[]{\label{subfig:i-m:post_optics}
    \includegraphics[width=0.42\columnwidth]{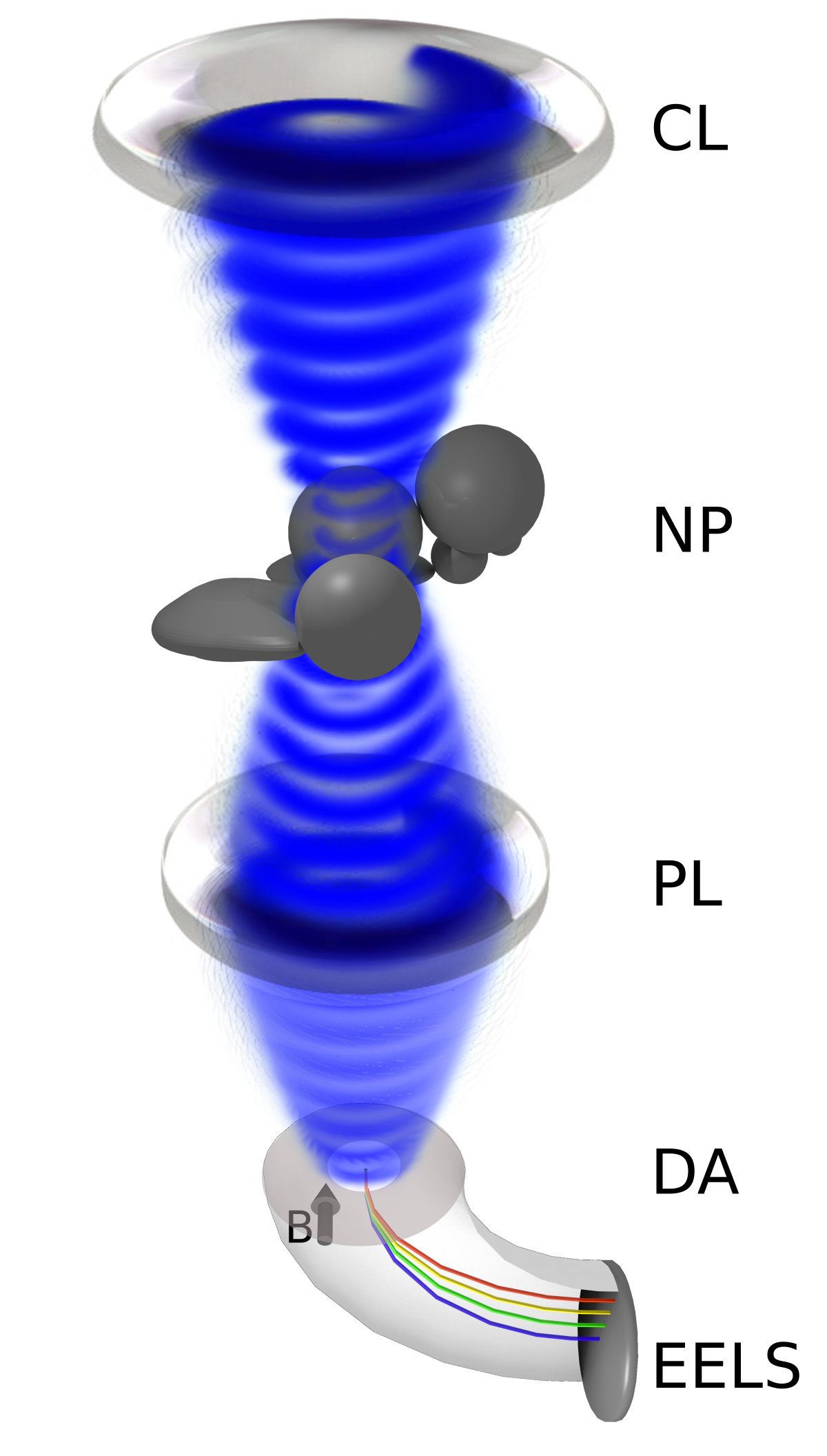} } \\
  \subfloat[]{\label{subfig:i-m:chiral}
    \includegraphics[width=0.47\columnwidth]{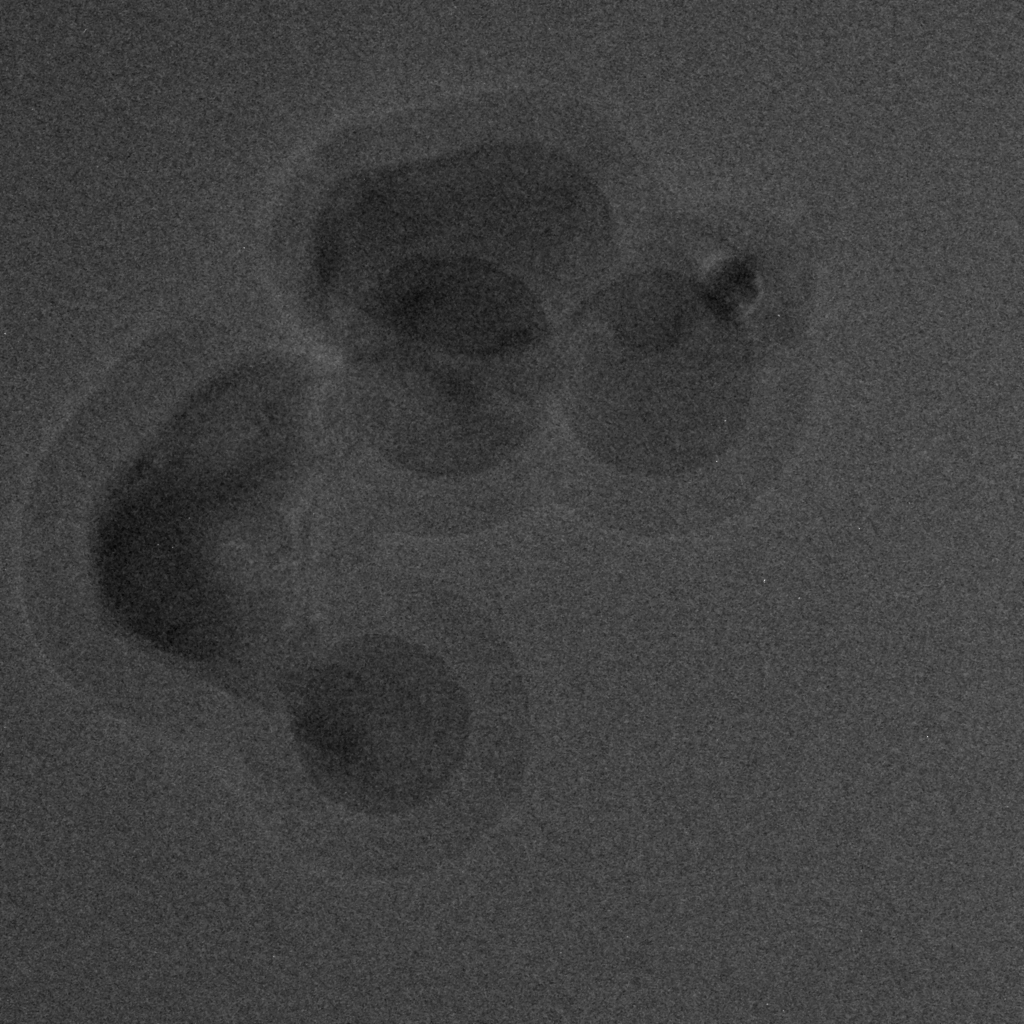} }
  \subfloat[]{\label{subfig:i-m:v_chiral}
    \includegraphics[width=0.48\columnwidth]{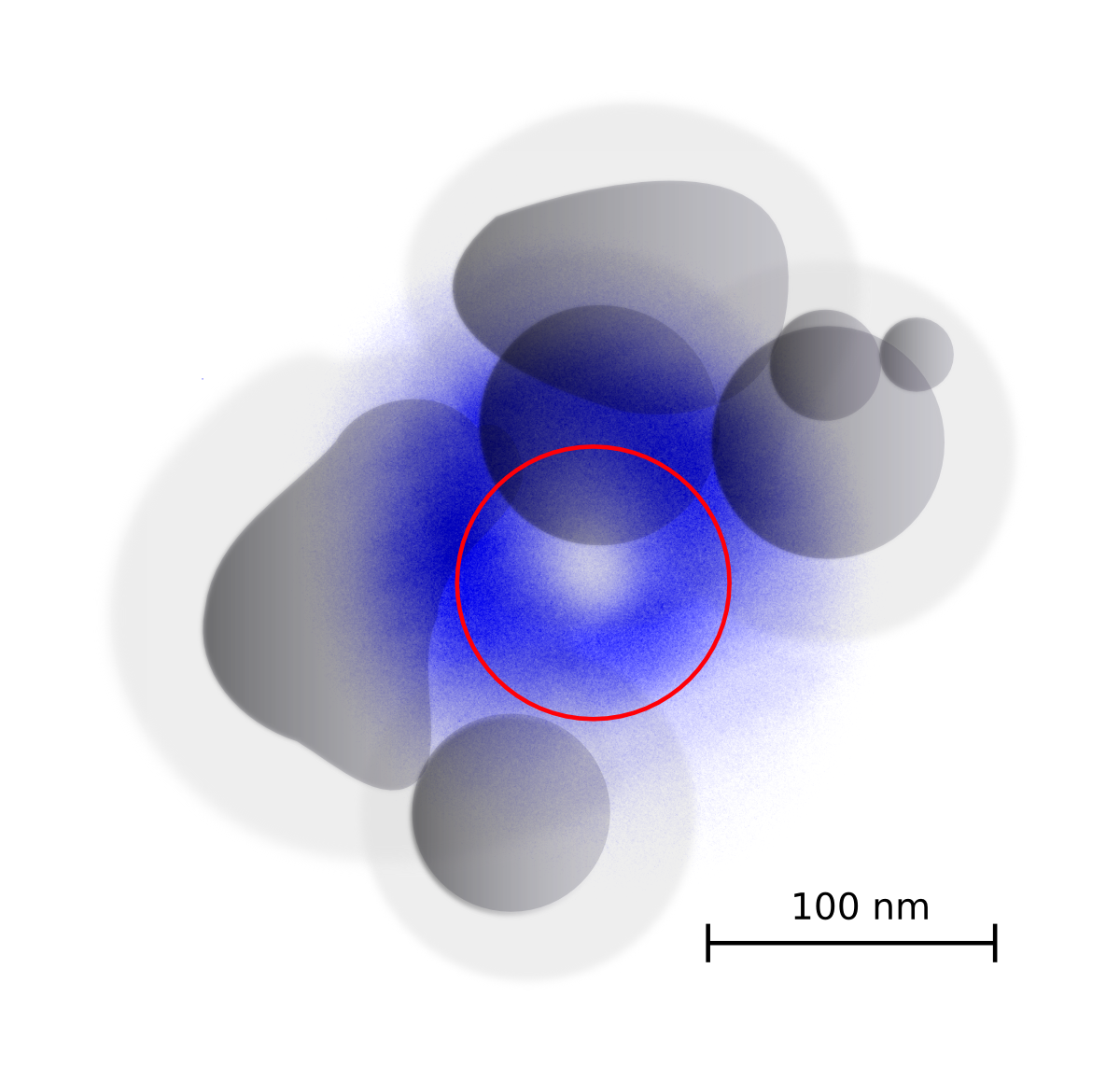} } \\
  \caption{
    (a) Simulated two-dimensional cross-sectional intensity (top) and radial intensity profile (bottom) of fully coherent beams with angular momentum quantum numbers $m=0$ (red), $m=1$ (blue) and $m=2$ (green) and other beam parameters held constant. Only the $m=0$ beam has non-zero amplitude at the center of the beams.
    (b) Schematic of post-specimen optical configuration. The condenser lens system (CL) focuses the incident vortex onto the chiral nanoparticle cluster in the specimen plane (NP); the projector lens system (PL) re-forms an image of the vortex-on-specimen at the round detector entrance aperture (DA). The electron energy-loss spectrometer (EELS), highly simplified in this schematic, separates the outgoing wave by energy to produce a spectrum. 
    (c) Transmission electron micrograph of chiral particle cluster under parallel illumination.
    (d) To-scale representation of chiral particles under vortex beam illumination. (grey) Aluminum particles (grey); (white) deposited hydrocarbon layer; (blue) $m_i = +1$ vortex beam; (red) actual spectrometer entrance aperture position used for one pair of electron energy-loss spectra.
    \label{fig:i-m}
  }
\end{figure}

The electron vortices employed in this experiment were produced in an FEI Titan transmission electron microscope (TEM) at 300 kV. We placed a nanofabricated forked diffraction grating in the condenser lens aperture of the microscope, as shown in Fig. \ref{subfig:setup:grating} \cite{harvey_efficient_2014}. We then focused the $+1$ ($-1$) diffracted order of the grating to a nanoscale probe in the plane of the sample to be analyzed; this beam carried $m_i = +1$ ($-1$) quanta of orbital helicity \cite{mcmorran_electron_2011}. We examined chiral clusters of sub-\unit[100]{nm}-diameter aluminum nanoparticles, shown under planar illumination in Figure \ref{subfig:i-m:chiral}. Aluminum nanoparticles were well-suited for study in our instrument, as the electronic states we excited are most likey surface plasmon resonances, and aluminum nanoparticles have documented surface plasmon resonances well into the UV (i.e. above \unit[3]{eV}) \cite{langhammer_localized_2008}; these resonances were easily resolvable on our instrument, which has a spread in incident electron energy on the order of \unit[1]{eV}.

A successful dichroism experiment depends on pre- and post-selection of angular momentum. In the case of optical and X-ray circular dichroism, in which the incident spin is fully transferred to the specimen when a photon is absorbed, a post-selection for a vacuum state means $m_f = 0$ and thus one quantum of angular momentum is always transferred to the specimen per photon. To measure electron helical dichroism, we must be more careful; the electron we send into a specimen exits with a spectrum of energies and orbital angular momenta. We seek to pick out the electron $m_f = 0 $ state. We compare the inital and final states in circular dichroism and electron helical dichroism in Table I\ref{t:dich}. Electron energy loss measurement devices are commercially available for electron microscopes, but orbital angular momentum measurement is not so well-developed. 
\begin{table}[h]
  \[  \begin{array}{l|l|r} \label{t:dich}
    \textrm{Particle used as probe} & \textrm{Initial probe state} & \textrm{Final state} \\
\hline 
\textrm{Circ. polarized photon} & \ket{\mathbf{k_i},m_s = \pm 1}  & \ket{\mathbf{0}} \\
\textrm{Electron vortex} & \ket{\mathbf{k_i},m_i}  & \ket{\mathbf{k_f},m_f} \\
\end{array} \]
\caption{In a simplified decription of circular dichroism, measurement of absorption of a photon probe $\ket{\mathbf{k_i}, m_s = \pm 1}$, i.e. post-selection for the vacuum state $\ket{\mathbf{0}}$, automatically gaurantees post-selection for a zero-angular momentum $m_s = 0$ final state as a zero-photon state carries no angular momentum. However, as massive electrons cannot annihilate to a vacuum state, we must carefully prepare the incident electron probe state $\ket{\mathbf{k_i},m_i}$ and post-select for the final electron state $\ket{\mathbf{k_f},m_f}$.}
\end{table}

As yet, no quantitative, non-interferometric orbital angular momentum measurement technique exists for electrons. Interferometric measurement techniques \cite{clark_quantitative_2014,guzzinati_measuring_2014} cannot measure the orbital angular momentum of the incoherent superposition of energy and orbital angular momentum eigenstates produced by inelastic scattering in a specimen \cite{schattschneider_mapping_2012}. Fortunately, we did not need a sophisticated orbital angular momentum measurement technique to select for the final state that corresponds to a transfer of one quantum of orbital angular momentum to the specimen.

Instead, we performed a simple post-selection for the zero-orbital angular momentum final state. We admitted only the center of an in-focus projection of the incident vortex beam on the specimen through an aperture, into the detector, as shown in Figure \ref{fig:i-m}. This post-selection weights more heavily the component of the outgoing wave that transferred all incident orbital angular momentum to the specimen \cite{harvey_electron_2013,harvey_characterization_2014}. We detail this post-selection in Section \ref{sect:ps} in the Supplemental Material \cite{supp}.

We sought to use this centered-aperture post-selection technique to control the angular momentum transfer to characterize the chirality of our aluminum nanoparticle cluster. We measured electron energy-loss spectrum pairs $J_+(E)$ for $m_i = +1$ and $J_-(E)$ for $m_i = -1$ and a more heavily weighted $m_f = 0$ final state with this post-selection. The peak in the dichroic electron energy-loss spectrum $J_{\textrm{d}} = J_+ - J_-$ produced is a measurement of the difference in differential scattering probability $\deriv{P_{\pm}}{E}$ for excitation of electronic states with opposite angular momentum:
\begin{equation}
  J_{\textrm{d}} \propto \deriv{P_+}{E} - \deriv{P_-}{E}.
\end{equation}
An extended explanation of helical dichroism measurement can be found in Section \ref{sect:dich} in the Supplemental Material \cite{supp}.

\begin{figure*}[tbh]
  \subfloat[]{\label{subfig:EELS:full}
    \includegraphics[width=0.47\textwidth]{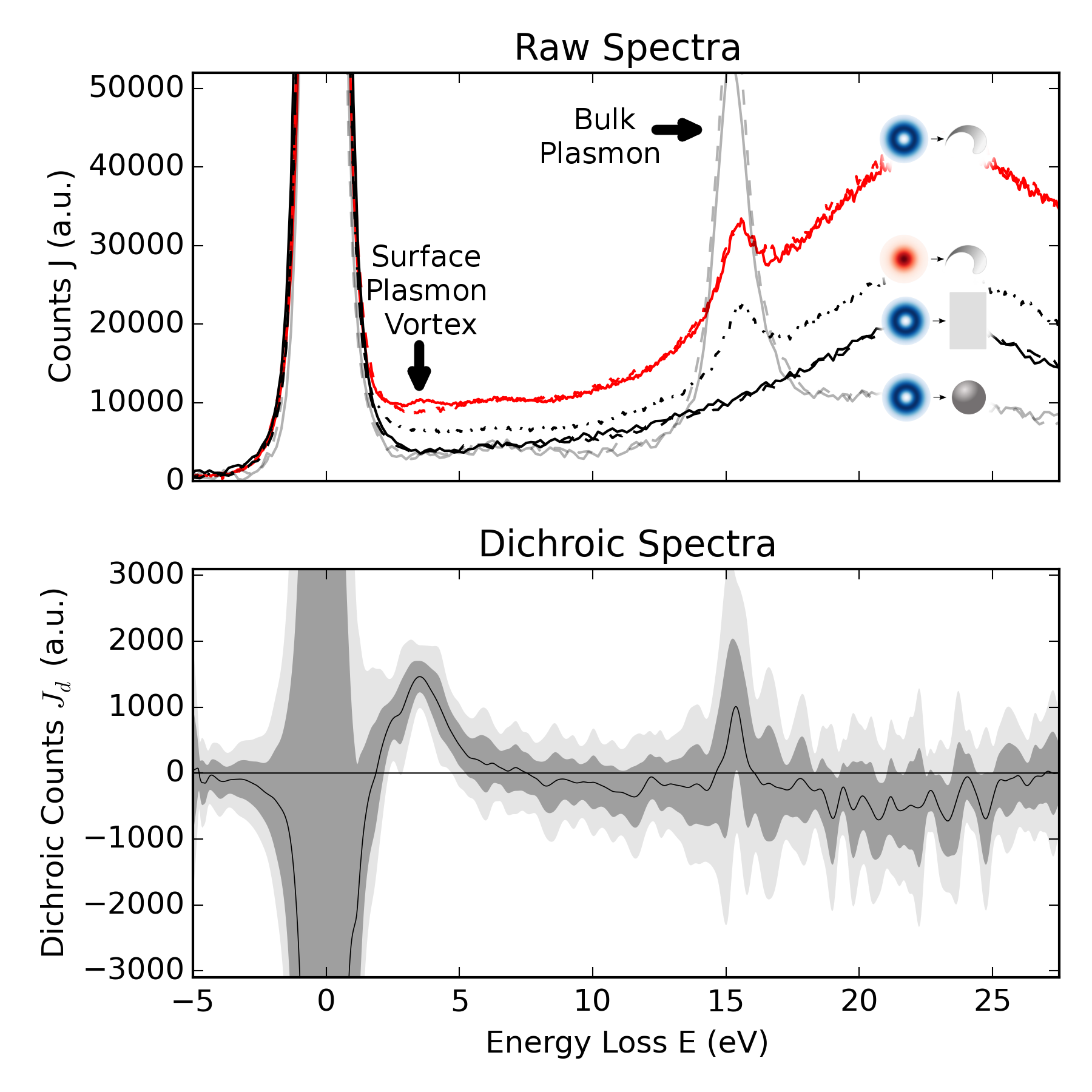}}
  \subfloat[]{\label{subfig:EELS:highres}
    \includegraphics[width=0.47\textwidth]{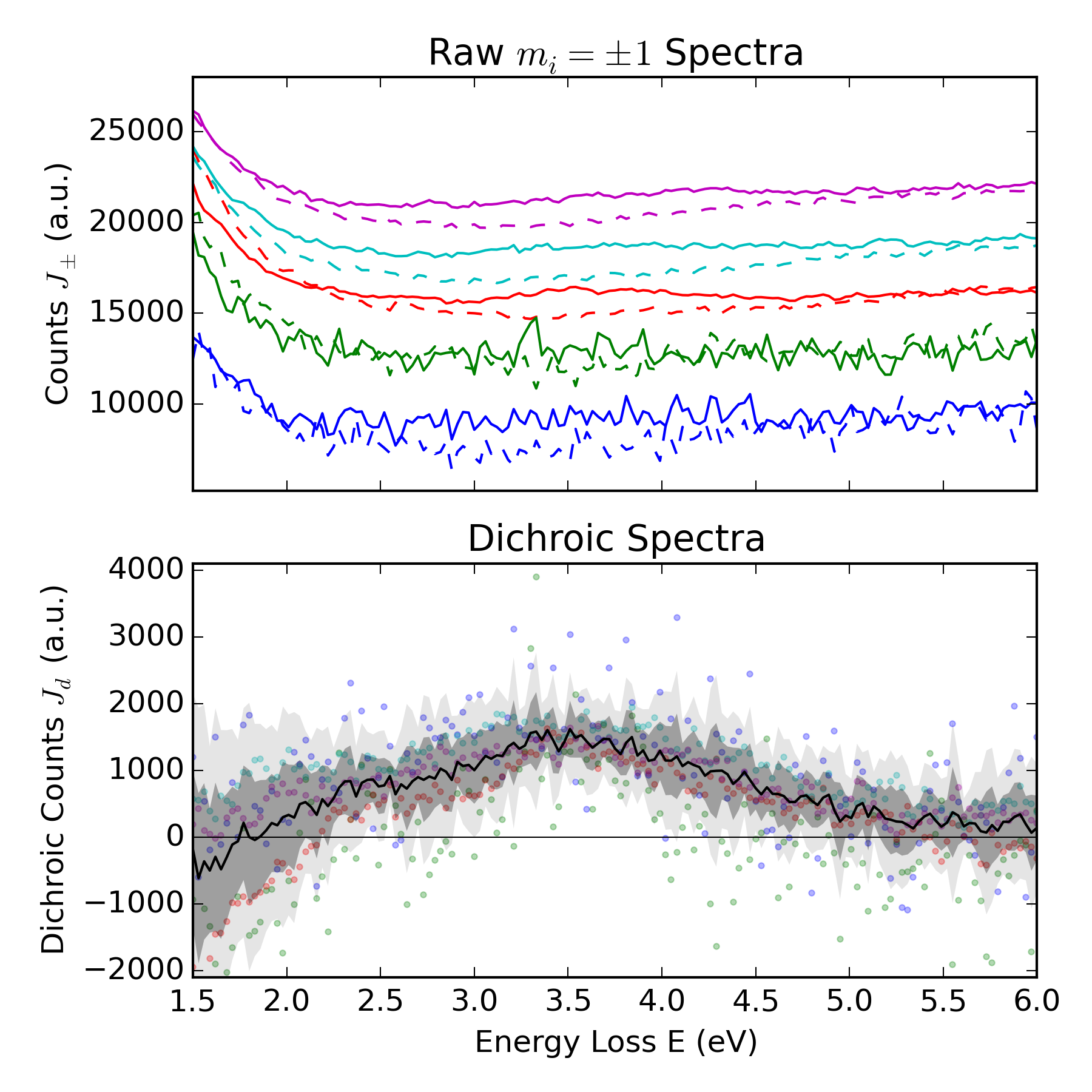}}\\
    \caption{
      (a) (\textbf{top}) Comparison of electron energy-loss spectra for several incident wavefunction and specimen permutations. Dichroism appears at $\unit[3.5\pm 0.8]{\textrm{eV}}$ in the energy-loss spectrum pair $J_+(E)$ (red, solid) and $J_-(E)$ (dashed) produced by interaction of vortices with the chiral cluster of Al nanoparticles on a $\textrm{Si}\textrm{N}_x$ substrate, shown in Fig. \ref{subfig:i-m:chiral}. Spectrum pairs produced by interaction with the $\textrm{Si}\textrm{N}_x$ substrate without nanoparticles (black) or a spherically symmetric nanoparticle (grey) show no significant differences. Energy loss from an $m_i=0$ planar wave (black, dash-dot) due to interaction with the same Al chiral cluster has no distinct peaks in the $\unit[3.5\pm 0.8]{\textrm{eV}}$ range.
      (\textbf{bottom}) The average of all dichroic spectra (black line) shows a significant dichroic peak at $\unit[3.5\pm 0.8]{\textrm{eV}}$. The 68\% (dark grey region) and 95\% (light grey region) confidence intervals are smoothed for clarity. The extreme width of the confidence intervals near zero energy loss illustrate that small variations in beam position cause large fluctuations in zero-loss intensity with our post-selection technique.  \\
      (b) (\textbf{top}) Electron energy-loss spectrum pairs $J_+$ for the interaction of a single-helix vortex beam with $m_i = +1$ (solid) and $J_-$ for $m_i = -1$ (dashed) with the chiral cluster. Spectra are offset for clarity, and blue and green spectra were recorded with a shorter exposure to control for the possibity of beam position fluctuation.
      (\textbf{bottom}) Dichroic spectra $J_d(E)$ (colored points) calculated by subtraction of $J_+$ spectra from $J_-$ spectra shown above; the average (black line) of all dichroic spectra deviates from zero dichroic counts with 68\% confidence (dark grey region) only in this energy range. In fact, the dichroic peak deviates from zero with 95\% confidence (light grey region) over a $\unit[1]{eV}$ range. \\ 
        \label{fig:EELS} }
\end{figure*}

Great care must be taken to produce a dichroism spectrum. A small mis-assignment of energies to peaks in any two consecutive energy loss spectra, such as our dichroic pairs, can produce apparently significant dichroism. The beam-specimen interaction recorded in the spectrometer can change in response to any small drift in the position of the specimen relative to the beam, caused by, for example, thermal expansion of the stage or fluctuations in temperature, pressure or stray magnetic or electric fields in the microscope room. We controlled for these potential issues in this experiment by intentionally re-aligning the beam, specimen and post-selection aperture and recording spectra over multiple timescales so that our measured uncertainty accurately reflects all relevant fluctuations in our microscope. 

We also controlled for the possbility that small differences in the wavefunctions of beams with opposite incident helicity could produce spurious dichroism. We would expect to see such an artifact on specimens without explicit chiral structure. However, We observed no significant dichroic peaks for vortex beams incident on a symmetric aluminum nanoparticle or a silicon nitride substrate (Fig. \ref{subfig:EELS:full}). 

Dichroic spectra on the chiral aluminum cluster in Figure \ref{subfig:EELS:highres} show one significant UV peak at $\unit[3.5 \pm 0.8]{\textrm{eV}}$. This dichroic peak likely corresponds to excitation of surface plasmon vortex states.
A surface plasmon vortex is a collective excitation of surface conduction electrons that carries orbital angular momentum, and therefore features the same azimuthal phase structure carried by an electron vortex \cite{ohno_study_2006}. Three observations support the identification of this peak as surface plasmon vortex excitation: 
\begin{enumerate}
  \item The majority of the post-selected wavefunction that enters the spectrometer does not directly pass through the aluminum particle, but rather passes near the surface, as shown in Figure \ref{subfig:i-m:v_chiral}.
  \item Isolated aluminum nanoparticles with diameters on the order of \unit[100]{nm} have dipolar plasmon resonances on the order of \unit[3]{eV} \cite{langhammer_localized_2008}; we should therefore expect a surface plasmon vortex mode on a chiral cluster of similarly-sized particles at the same order of magnitude. 
  \item Localized surface plasmon resonances are strongly sensitive to geometery, and in particular, surface plasmon vortex excitation is sensitive to structural chirality \cite{ohno_study_2006,gorodetski_observation_2008}. Not every chiral cluster for which we recorded spectra showed a strong dichroic peak, which suggests a strong dependence on particle size, shape and cluster configuration. We discuss further observations of dichroism in section \ref{sect:supp} in the Supplemental Material \cite{supp}. 
\end{enumerate}

We therefore conclude that the observed dichroic peak on the chiral cluster likely corresponds to excitation of a surface plasmon vortex. There may be other significant contributions under \unit[1]{eV}, or less signficant peaks of the opposite sign, that might be identifiable in an electron helical dichroism experiment conducted on an instrument with better energy resolution. A Nion HERMES instrument will soon be equipped to probe dichroism at much lower energies \cite{krivanek_toward_2014}.

We have demonstrated electron helical dichroism on a chiral cluster of aluminum nanoparticles. We inferred that the dichroic peak represents a difference in the density of plasmon vortex states for this cluster. With further development, electron helical dichroism may be valuable for high-spatial resolution measurement of chirality over a large range of energies. 

In particular, fully-developed chirality measurement with electrons demands a theoretically predicted and experimentally validated quantitative description of the relationship between the geometric chirality of a material and the magnitude of dichroism. If further work can shed light on this connection, electron helical dichroism may be employed to map out the chirality of single biomolecules and nanostructures with high spatial resolution over infrared, visible, ultraviolet and X-ray energies.

As structural determination of non-crystalline biomolecules at near-atomic resolution remains challenging \cite{bai_ribosome_2013,zhou_towards_2008}, structural biology stands to benefit from a new tool to measure chirality and infer structure in lieu of perfect atom-by-atom imaging. Furthermore, the ability to map surface plasmons with orbital angular momentum in the transmission electron microscope could illuminate future plasmon vortex-based device engineering. Lastly, the tools developed for this electron helical dichroism measurement may be transferrable to the advancement of other applications for dichroism measurement with electron vortices.

\begin{acknowledgments}
The authors wish to thank Josh Razink of the CAMCOR High Resolution and Analytical facility at University of Oregon for instrument support, Peter Ercius and Colin Ophus of the National Center for Electron Microscopy for critical conversations and technique advice, and Miriam Deutsch of the University of Oregon Department of Physics for informative conversations regarding surface plasmonics and manuscript advice. We gratefully acknowledge the use of CAMCOR facilities, which have been purchased with a combination of federal and state funding. This work was supported by the U.S. Department of Energy, Office of Science, Basic Energy Sciences, under Award DE-SC0010466. 
\end{acknowledgments}

\bibliographystyle{apsrev4-1}
\bibliography{bibtex}{}

\begin{appendices}
\widetext
\clearpage

\begin{center}
\textbf{\large Supplemental Material for ``Demonstration of electron helical dichroism as a local probe of chirality''}
\end{center}

\section{\label{sect:ps}Post-selection} 

The post-selection technique we use in this work is based upon the spatial distribution of intensity caused by the singularity of vortex beams, which we developed in more detail in past work \cite{harvey_electron_2013,harvey_characterization_2014}.

The peak intensity of a vortex beam forms a ring (Fig. \ref{subfig:setup:beam}). Inside this ring, the intensity of a fully coherent vortex drops to zero, because the azimuthal phase term of a vortex beam, $e^{i m \phi}$, is singular at the center of the beam. The orbital angular momentum of a vortex beam is conserved in free space; in other words, the hole in the center of a vortex is stable under free-space propagation \cite{bliokh_semiclassical_2007}. Equivalently, because transverse variations in phase grow infinitely large towards the center of the beam, the wavefunction of a vortex is diffraction-limited at its center and must develop a hole upon far-field propagation. A gaussian beam with no angular momentum, however, has no phase singularity and the peak intensity is at the center of the beam (Fig. \ref{subfig:i-m:superposition}). We can utilize this difference in the position of peak intensity to perform a simple post-selection for the component of an outgoing wave that transferred all incident orbital angular momentum to the specimen.

If an incident electron vortex beam with $m_i = \pm 1$ quantum of angular momentum along the propagation axis transfers that angular momentum via interaction with a specimen, the full-OAM-transfer $\Delta m \equiv m_i - m_f = \pm 1$ scattered component of the beam that has $m_f = 0$ will cause an increase in the intensity of the center of the beam. However, if the opposite angular momentum transfer occurs upon interaction, with $\Delta m = \mp 1$, the scattered component of the beam that corresponds to this interaction has $m_f = \pm 2$ and retains the singularity at the center of the beam. The distinct radial profiles of $m = 0$, $m=1$ and $m=2$ beams with a common radial quantum number are illustrated in Figure \ref{subfig:i-m:superposition}. An aperture with a radius equal to the radius of peak intensity of the $m=1$ vortex would allow 63\% of the intensity of the $m=0$ beam to pass, but only 8\% of the $m=2$ beam. Therefore, a spatial post-selection can serve as a qualitative measurement of orbital angular momentum transfer: any measurement that weights intensity at the center of the outgoing wave higher than intensity at larger radii will weight the scattering amplitude for the full-transfer $\Delta m = \pm 1$ transition more heavily than the $\Delta m = \mp 1$ transition.

\section{\label{sect:dich}Electron Helical Dichroism: An Introduction} 
There are 3 steps necessary for an electron helical dichroism experiment. First, one records the electron energy-loss spectrum $J_{\pm}(E)$ for the orbital angular momentum-transfer interaction (in this experiment, the $m_f=0$ final state) for both positive and negative-orbital angular momentum initial states. Then, since the total intensity produced by the post-selection scheme described in Section \ref{sect:ps} is highly sensitive to fluctuations in beam position, one must normalize the spectra so that non-chiral peaks are equally large. Lastly, one subtracts the negative-orbital angular momentum-transfer spectrum $J_-$ from the positive-orbital angular momentum-transfer spectrum $J_+$ to produce an EHD spectrum. We used initial orbital angular momentum states $m_i = \pm 1$. As all $m_f = m_i$ signal will cancel in the dichroic spectrum, since the probability for $\Delta m = 0$ inelastic scattering does not depend on the sign of $m_i$, and as higher-order orbital angular momentum transfer has negligible measured intensity due to our post-selection, the dichroic spectrum $J_{\textrm{d}}$  is a measurement of the difference in differential scattering probability $\deriv{P_{\pm}}{E}$ for excitation of surface plasmon vortices of opposite angular momentum:
\begin{equation}
  J_{\textrm{d}} = J_+(E)-J_-(E) \propto \deriv{P_+}{E} - \deriv{P_-}{E}
\end{equation}

As current orbital angular momentum measurement techniques for electrons cannot post-select for a single final orbital angular momentum state from an incoherent superposition of states with varying energy, we use the spectrometer entrance aperture to post-select for the $m_f = 0$ final state. 

We measured the electron energy-loss spectrum of the central scattered portion of an electron vortex with $m_i = \pm 1$ orbital angular momentum incident on a chiral distribution of aluminum nanoparticles. To do so, we centered the particle cluster and electron vortex over the entrance aperture of the electron energy-loss spectrometer in order to increase the ratio of the intensity of the central scattered $m_f=0$ portion of the beam relative to the unscattered $m_f=m_i=\pm 1$ beam that passes into the spectrometer. Electron energy-loss spectra recorded this way are shown in Figure \ref{subfig:EELS:full}. We then subtracted $J_-$ spectra recorded for the $m_i = -1 $ incident beam from the $m_i = +1 $-incident $J_+$ spectra. These dichroic spectra are shown in Figure \ref{subfig:EELS:highres}. As the alignment of the incoming vortex, particle cluster and entrance aperture are crucial for good post-selection of the zero-orbital angular momentum component of the outgoing wave, we recorded $J_+$ and $J-$ spectrum pairs with five independent alignments of the beam, specimen and aperture. Indeed, we see significant variation in the non-chiral plasmonic spectrum among the five spectrum pairs shown in Figure \ref{subfig:EELS:highres}, but a consistent difference between $J_+$ and $J_-$.

\section{\label{sect:supp} Supplemental Observation of Dichroism}

\begin{figure}
  \subfloat[]{\label{supp:subfig:EELS}
    \includegraphics[width=0.62\columnwidth]{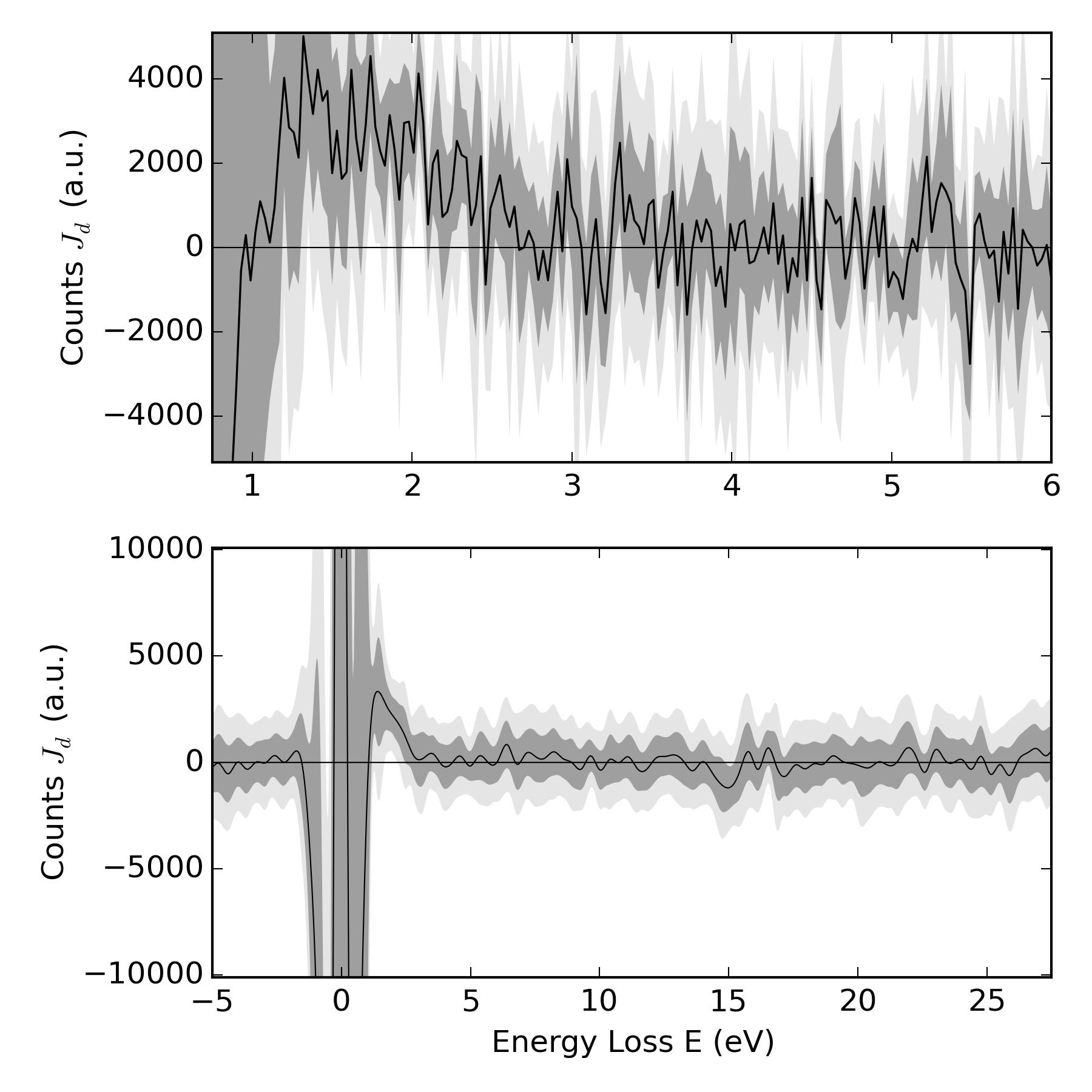} }
  \subfloat[]{\label{supp:subfig:particles}
    \includegraphics[width=0.38\columnwidth]{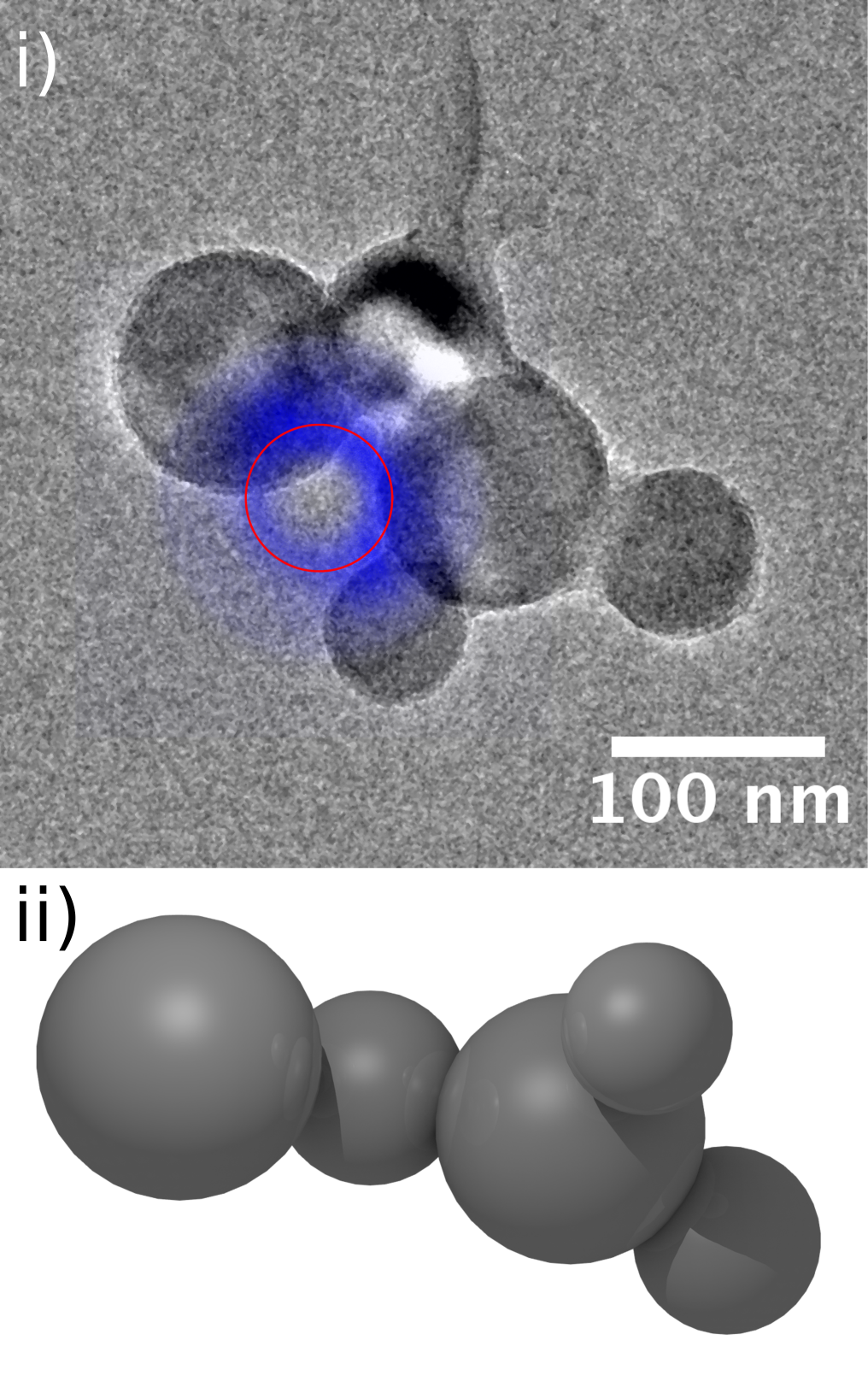} }
  \caption{
    (a) Second observation of dichroic electron energy-loss spectra on a chiral particle cluster.
    (b) (i) Micrograph of particle cluster with orientation, beam positon (blue) and aperture position (red) used in acquisition of dichroic spectra shown in (a). (ii) Three-dimensional model of the chiral particle cluster, reconstructed by tomographic tilt series.
        \label{supp:fig:EELS} }
\end{figure}

As a further control for the possibility of misalignment-induced spurious dichroism, we performed an electron helical dichroism experiment on a second chiral cluster. We observed a similar peak at $\unit[1.8\pm 0.5]{\textrm{eV}}$, and no other significant peaks, as shown in Figure \ref{supp:subfig:EELS}. We also reconstructured a three-dimensional model of the particle cluster by tomographic tilt series, shown in Figure \ref{supp:subfig:particles}.

We recorded spectra on several other chiral clusters which showed a barely-significant or no significant dichroic peak in the same energy range. We speculate that this dearth of dichroism on some clearly chiral clusters results from the strong dependence of surface plasmon resonances on particle size, shape and cluster structure: we may have incidentally gathered spectra from a location on the cluster with a relatively weak plasmon resonance. Further development of electron helical dichroism may allow for rapid spectrum acqusition across all positions on a chiral structure.

\end{appendices}

\end{document}